\documentclass[aps, showpacs, showkeys,nofootinbib,floatfix ,twocolumn]{revtex4}

\usepackage{amssymb}
\usepackage{amsmath}
\usepackage{graphicx}

\begin{document}

\title{Viscous Generalized Chaplygin Gas as a Unified Dark Fluid: Including Perturbation of Bulk Viscosity}

\author{Wei Li$^{1,2}$}
\author{Lixin Xu$^{1}$\footnote{corresponding author: lxxu@dlut.edu.cn}}


\affiliation{$^{1}$Institute of Theoretical Physics, School of Physics \& Optoelectronic Technology, Dalian
University of Technology, Dalian, 116024, P.R.China}
\affiliation{$^{2}$Department of Physics, College of Mathematics and Physics, Bohai University, Jinzhou,
121013, China}

\begin{abstract}
In this paper, we continue our previous work of studying viscous generalized Chaplygin gas (VGCG)
as a unified dark fluid but including the bulk viscosity perturbation.  By using the currently available cosmic observational
data from SNLS3, BAO, HST and recently released Planck, we gain the constraint on bulk viscosity coefficient: $\zeta_0=0.0000138_{-    0.0000105-    0.0000138-    0.0000138}^{+    0.00000614+    0.0000145+    0.0000212}$ in $1, 2, 3\sigma$ regions respectively via Markov Chain Monte Carlo method. The result shows that when considering perturbation of bulk viscosity, the currently cosmic observations favor a smaller bulk viscosity coefficient.
\end{abstract}

\maketitle

\section{Introduction}
Several astronomical observations such as SN Ia \cite{1}, CMB  \cite{2}, WL \cite{3}, etc. powerfully indicates that in the present, the overwhelming majority of cosmological total energy content is a dark sector which takes charge of the acceleration of our universe. This dark sector is generally assumed owning two different components: dark energy and dark matter. To investigate this dark sector, many cosmological models are built basing on the cosmological principle validity and the assumption of an idealized perfect fluid, which means that all components of the matter-energy in our universe are considered as perfect fluid without any viscosity. The most competitive model of dark energy is a cosmological constant model. But researches have shown that constant dark energy models are not well confirmed by both observations and theoretical considerations \cite{21, 22}. One of the alternatives to the cosmological constant is to describe dark matter and dark energy within a unified dark fluid model. For all we know, the Chaplygin gas \cite{4}\cite{5,6} was firstly presented along this line. However, the unified Chaplygin gas type models forecasted instabilities or mighty oscillations of small scale in the matter power spectrum, which disagrees with the observational data \cite{7}. This problem \cite{8,16} may be alleviated or even avoided by the non-adiabatic perturbations.  A reasonable possibility is to allow the gCg to have non-adiabatic perturbations, which is a natural assumption since it is not a pressureless fluid actually. An attempt in this direction has already been performed in \cite{18} and \cite{19}.  Furthermore, in the recently years, more and more cosmological observations suggest that our universe is permeated by imperfect fluid, in which the negative pressure, as was argued in \cite{16,17}, an effective pressure including bulk viscosity can play the role of an agent that drives the present acceleration of universe.

The viscous generalized Chaplygin gas (hereafter referred to as VGCG) is a widely studied model among those proposed to describe the observed accelerated expansion of the universe. In contrast to many models describing dark energy alone, the VGCG gives a unified description of dark matter and dark energy, enrolling itself in the class of so-called unified dark fluid (UDF) cosmological models see e.g. \cite{9,10,11}. A common characteristic of these papers is that only the impact of bulk viscosity on the background expansion of the universe is studied without considering perturbation of bulk viscosity. However, the perturbation analysis of the viscous cosmological models is crucially important to the evolution of cosmology. The different mentioned approaches imply a generally different dynamics at the perturbative level. Therefore, it is interesting to study the behaviour of the VGCG under perturbations.

In the present paper, we study only scalar perturbations following the notation of \cite{14}. we will modify the pressure through Eckart¡¯s expression \cite{23} $p_{v}=-\zeta u^{\mu}_{;\mu}$, where bulk viscosity coefficient $\zeta$ is a non-negative quantity, and the fluid-expansion scalar $u^{\mu}_{;\mu}$ is reduces to $3H$ in the isotropic and homogeneous universe, where $H=\frac{\dot{a}}{a}$ is the Hubble parameter.
As a continuation of our previous work \cite{15}, here we investigate VGCG model by including bulk viscosity perturbation.

The structure of this paper is organized as follows. In the next
section, we briefly introduce some basic equations of viscous generalized Chaplygin gas model. The derivation of evolution equations for density perturbation and velocity perturbations are presented
in the third section. Then in the forth section, by using the MCMC
method, we perform a global fitting
to the currently observational data and analyze the constraint results. The discussion
and conclusion are given in the final section.

\section{basic equations of viscous generalized chaplygin gas model}

In an isotropic and homogeneous universe, we consider the standard
Friedmann-Robertson-Walker metric,
\begin{equation}
ds^{2}=-dt^{2}+a^{2}(t)\left[\frac{1}{1-kr^{2}}dr^{2}+r^{2}(d\theta^{2}+\sin^{2}\theta
d\phi^{2})\right].
\end{equation}
For the sake of simplicity, we choose the flat geometry $k=0$, which
is also favored by the update result of the cosmic background
radiation measurement. The general stress-energy-momentum tensor is
\begin{equation}
T_{\mu\nu}=(\rho+p)U_{\mu}U_{\nu}+pg_{\mu\nu}.
\end{equation}
To consider the effect of bulk viscosity, we modify the pressure
only by redefining the effective pressure $p_{eff}$, according to
$p_{eff}=p+p_{v}=p-3H\zeta$, we re-write the viscous energy-momentum tensor
\cite{Jean} as:
\begin{eqnarray}
\label{eq:Tmunu_viscous_2} T_{\rm viscous}^{\mu\nu} & = & \rho
U^{\mu}U^{\nu} + \left(p -3\zeta
\frac{\dot{a}}{a}\right)\left(g^{\mu\nu}
+ U^{\mu}U^{\nu}\right) \nonumber \\
                                                 & \equiv & p_{\rm eff}g^{\mu\nu} + (p_{\rm
eff} + \rho)U^{\mu}U^{\nu}.
\end{eqnarray}

From the equation above, we see that the effect of bulk viscosity is
to change the pressure $p$ to an effective pressure $p_{\rm eff} = p
-3\zeta \dot{a}/a$.  The physical interpretation is clear that a
viscous pressure can play the role of an agent that drives the
present acceleration of the universe. Note that the possibility of a
viscosity dominated late epoch of the universe with accelerated
expansion was already mentioned by Padmanabhan and Chitre in
\cite{Space Sci}.

Using the GCG equation of state $p_{g}=-A/\rho_{g}^{\alpha}$, which
yields an analytically solvable cosmological dynamics if the
universe is GCG dominated, we obtain the equation of state (EoS) of
viscous GCG (VGCG) model is given in the form of
\begin{equation}
p_{VGCG}=-A/\rho^{\alpha}_{VGCG}-3H\zeta\\,
\end{equation}
this EoS includes the GCG model as its special case when $\zeta=0$;
when $\zeta\neq0$, for the normal form
$\zeta=\zeta_{0}\rho_{VGCG}^{\frac{1}{2}
}$, we have the equation of
state {EoS}
\begin{equation}
p_{VGCG}=-A/\rho^{\alpha}_{VGCG}-\sqrt{3}\zeta_{0}\rho_{VGCG},
\end{equation}
where $A$, $\zeta_{0}$and $\alpha$ are model parameters. Applying
the energy conservation of VGCG, one can deduce its energy density
as
\begin{eqnarray}
\rho_{VGCG}&=&\rho_{VGCG0}[\frac{B_{s}}{1-\sqrt{3}\zeta_{0}}+(1-\frac{B_{s}}{1-\sqrt{3}\zeta_{0}})\nonumber
\\&\times&a^{-3(1+\alpha)(1-\sqrt{3}\zeta_{0})}]^{\frac{1}{1+\alpha}},\label{eq:mcg}
\end{eqnarray}
where $B_{s}=A/\rho^{1+\alpha}_{VGCG0}$, $\alpha$ and $\zeta_{0}$ are
model parameters. Form Eq.(\ref{eq:mcg}), one can find that $0\le
B_s \le 1$ and $\zeta_{0}<\frac{1}{\sqrt{3}}$ are demanded to keep
the positivity of energy density. If $\alpha=0$ and $\zeta_{0}=0$ in
Eq.(\ref{eq:mcg}), the standard $\Lambda$CDM model is recovered.
Taking VGCG as a unified component, one has the Friedmann equation
\begin{eqnarray}
H^{2}&=&H^{2}_{0}\{(1-\Omega_{b}-\Omega_{r}-\Omega_{k})[\frac{B_{s}}{1-\sqrt{3}\zeta_{0}}\nonumber\\
&&+(1-\frac{B_{s}}{1-\sqrt{3}\zeta_{0}})a^{-3(1+\alpha)(1-\sqrt{3}\zeta_{0})}]^{\frac{1}{1+\alpha}}\nonumber\\
&&+\left.\Omega_{b}a^{-3}+\Omega_{r}a^{-4}+\Omega_{k}a^{-2}\right\},
\end{eqnarray}
where $H$ is the Hubble parameter with its current value
$H_{0}=100h\text{km s}^{-1}\text{Mpc}^{-1}$, and $\Omega_{i}$
($i=b,r,k$) are dimensionless energy parameters of baryon, radiation
and effective curvature density respectively. In this paper, we only
consider the spatially flat FRW universe.

Here, we treat
VGCG as a unified dark fluid which interacts with the remaining
matter purely through gravity. With assumption of pure adiabatic
contribution to the perturbations, the adiabatic sound speed for
VGCG is
\begin{equation}
c^2_{s,ad}=\frac{\dot{p}_{VGCG}}{\dot{\rho}_{VGCG}}=-\alpha
w_{eff}-\sqrt{3}\zeta_{0},\label{eq:cs2}
\end{equation}
where $w_{eff}$ is the EoS of VGCG in the form of
\begin{equation}
w_{eff}=-\frac{B_{s}}{B_{s}+(1-B_{s})a^{-3(1+\alpha)}}-\sqrt{3}\zeta_{0}.
\end{equation}
From the above equation, one can find that in order to protect the
 sound of speed from negativity , $\alpha\ge 0$ is required because
of the non-positive values of $w_{eff}$.

We studied the perturbation evolution equations of VGCG in order to
research the effects on CMB anisotropic power spectrum.
In the synchronous gauge, using the conservation of energy-momentum
tensor ${T_{(viscous)}}^{\mu}_{\nu;\mu}=0$, one has the perturbation
equations of density contrast and velocity divergence for VGCG
\begin{eqnarray}
\dot{\delta}_{VGCG}&=&-(1+w_{eff})(\theta_{VGCG}+\frac{\dot{h}}{2})\nonumber\\
&-&3\mathcal{H}(c^{2}_{s}-w_{eff})\delta_{VGCG},\\
\dot{\theta}_{VGCG}&=&-\mathcal{H}(1-3c^{2}_{s})\theta_{VGCG}+\frac{c^{2}_{s}}{1+w_{eff}}k^{2}\delta_{VGCG}\nonumber\\
&-&k^{2}\sigma_{VGCG},
\end{eqnarray}
following the notation of Ma and Bertschinger \cite{ref:MB}. For the
perturbation theory in gauge ready formalism, please see
\cite{ref:Hwang}. The shear perturbation $\sigma_{VGCG}=0$ is
assumed and the adiabatic initial conditions are adopted in our
calculation. When the EoS of a pure barotropic fluid is negative, it
has an imaginary adiabatic sound speed which causes instability of
the perturbations , for example the $w=constant$ quintessence dark
energy model. The way to overcome this problem is to allow an
entropy perturbation and to assume a positive or null effective
speed of sound, which we will give a detailed study in the following.

\section{perturbation equations}
\subsection{perturbed Metric and Energy-momentum Tensor}
Scalar perturbations of the flat FRW metric are given in the
following form
\begin{eqnarray}
ds^{2}&=&a^{2}\{-(1+2\phi)d\tau^{2}+2\partial_{i}Bd\tau
dx^{i}\notag\\
&&+[(1-2\psi)\delta_{ij}+2\partial_{i}\partial_{j}E]dx^{i}dx^{j}\},
\end{eqnarray}
where $a$ is the scale factor, $\tau$ is the conformal time, $x_{i}$
are the spatial coordinates and $\phi$ and $\psi$ are the metric
perturbations.
The background four-velocity is $\bar{u}^{\mu}=a^{-1}(1,0,0,0)$,
which can be derived as follows,
\begin{equation}
\bar{u}^{\mu}=-\frac{d\tau}{ds}=-\frac{1}{a}\frac{dt}{ds}=\frac{1}{a}\delta^{\mu}_{0}.
\end{equation}
The spatial part is the perturbation, we can set it as
$\partial^{i}v$ for scalar perturbation only. Then using the
equality $g_{\mu\nu}u^{\mu}u^{\nu}=-1$, one has
\begin{equation}
u^{0}=\pm\frac{1}{\sqrt{{g}_{00}}}=+\frac{1}{a}(1-\phi),
\end{equation}
so one has the four-velocity of the fluid
\begin{eqnarray}
u^{\mu}&=&a^{-1}(1-\phi,~\partial^{i}v),\notag\\
u_{\mu}&=&g_{\mu\nu}u^{\nu}=a(-1-\phi,~\partial_{i}[v+B]),
\end{eqnarray}
where $v$ is the peculiar velocity potential. The local volume
expansion rate is $\theta=\vec{\nabla}\cdot\vec{v}$. Then one has
the expansion rate $\theta=-k^{2}(v+B)$ for the fluid.
 Let $u^{\mu}$ as the energy-frame four velocity (zero momentum flux relative to
 $u^{\mu}$). The energy density is its eigenvalue of this
 four-velocity, i.e., $T^{\mu}_{\nu}u^{\mu}=-\rho
 u^{\mu}$. The energy-momentum tensor can be written as
\begin{equation}
T^{\mu}_{\nu}=(\rho+p_{eff})u^{\mu}u_{\nu}+p_{eff}\delta^{\mu}_{\nu},
\end{equation}
where $\rho=\bar{\rho}+\delta\rho$ and $p=\bar{p}+\delta p$. The
effective pressure $p_{eff}$ is given as
\begin{eqnarray}
p_{eff}=p-\zeta\overline{(\nabla_{\gamma}u^{\gamma})}
=p-\frac{3\mathcal{H}}{a}\zeta.
\end{eqnarray}

The general energy-momentum tensor is
\begin{equation}
T^{0}_{0}=-\bar{\rho}-\delta\rho,
\end{equation}
\begin{equation}
T^{0}_{i}=(\bar{\rho}+\overline{{p_{eff}}})\partial_{i}(v+B)=(\bar{\rho}+\overline{{p_{eff}}})(v_{i}+B_{i}),
\end{equation}
\begin{equation}
T^{i}_{0}=-(\bar{\rho}+\overline{{p_{eff}}})v^{i},
\end{equation}
\begin{equation}
T^{i}_{j}=\overline{{p_{eff}}}\delta^{i}_{j}+\delta
p_{eff}\delta^{i}_{j}.
\end{equation}
Then one has the background energy-momentum tensor
\begin{equation}
\bar{T^{0}_{0}}=-\bar\rho,~~~~~\bar{T^{0}_{i}}=0,~~~~~\bar{T^{i}_{0}}=0,~~~~~\bar{T^{i}_{j}}=\overline{{p_{eff}}}\delta^{i}_{j}.
\end{equation}
Thus perturbed energy-momentum tensor can be written as
\begin{eqnarray}
\delta T^{0}_{0}&=&-\delta\rho,~~~\delta
T^{0}_{i}=(\bar{\rho}+\overline{{p_{eff}}})(v_{i}+B_{i}),\notag\\
 \delta
T^{i}_{0}&=&-(\bar{\rho}+\overline{{p_{eff}}})v^{i},~~~ \delta
T^{i}_{j}=\delta p_{eff}\delta^{i}_{j}.
\end{eqnarray}

\subsection{Calculation of Christoffel symbols}
The formula of the Christoffel symbols is given
\begin{equation}
\Gamma^{\mu}_{\alpha\beta}=\frac{1}{2}g^{\mu\nu}(g_{\alpha\nu,\beta}+g_{\beta\nu,\alpha}-g_{\alpha\beta,\nu}),
\end{equation}
where "$,$" stand for derivative, Greek letters
$\mu,\nu,\alpha,\beta$ take the values 0,1,2,3. In the following,
the prime "$\prime$" stand for derivative with respect to the
conformal time $\tau$.
So one has the following equations
\begin{equation}
\Gamma^{0}_{00}=\mathcal{H}+\phi^{\prime},
\end{equation}
\begin{equation}
\Gamma^{0}_{0i}=\phi_{i}+\mathcal{H}B_{i},
\end{equation}
\begin{equation}
\Gamma^{0}_{ij}=\mathcal{H}\delta_{ij}-[\psi^{\prime}+2\mathcal{H}(\psi+\phi)]\delta_{ij}+\partial_{i}\partial_{j}(E^{\prime}-B+2\mathcal{H}E),
\end{equation}
\begin{equation}
\Gamma^{i}_{00}=\partial^{i}(\phi+B^{\prime}+\mathcal{H}B),
\end{equation}
\begin{equation}
\Gamma^{i}_{j0}=\mathcal{H}\delta^{i}_{j}-\psi^{\prime}\delta^{i}_{j}+\partial_{j}\partial^{i}E^{\prime},
\end{equation}
\begin{equation}
\Gamma^{i}_{jk}=-\mathcal{H}\partial^{i}B\delta_{jk}+\delta_{jk}\partial^{i}\psi-\delta^{i}_{j}\partial_{k}\psi-\delta^{i}_{k}\partial_{j}\psi+\partial_{j}\partial_{k}\partial^{i}E.
\end{equation}
So, the nonzero Christoffel symbols are shown in the following, the
background items are
\begin{eqnarray}
\bar{\Gamma}^{0}_{00}=\mathcal{H},~~~~~\bar{\Gamma}^{0}_{ij}=\mathcal{H}\delta_{ij},~~~~~\bar{\Gamma}^{i}_{j0}=\mathcal{H}\delta^{i}_{j},
\end{eqnarray}
the perturbed items are
\begin{eqnarray}
\delta{\Gamma}^{0}_{00}&=&\phi^{\prime},~~~~~\delta{\Gamma}^{i}_{00}=\partial^{i}(\phi+B^{\prime}+\mathcal{H}B),\notag\\
\delta{\Gamma}^{0}_{0i}&=&\partial_{i}\phi+\mathcal{H}\partial_{i}B,
\delta{\Gamma}^{i}_{j0}=-\psi^{\prime}\delta^{i}_{j}+\partial_{j}\partial^{i}E^{\prime},\notag \\
\delta{\Gamma}^{0}_{ij}&=&-[\psi^{\prime}+2\mathcal{H}(\phi+\psi)]\delta_{ij}+\partial_{i}\partial_{j}[E^{\prime}+2\mathcal{H}E-B],\notag\\
\delta{\Gamma}^{i}_{jk}&=&-\mathcal{H}\partial^{i}B\delta_{jk}+\delta_{jk}\partial^{i}\psi-\delta^{i}_{j}\partial_{k}\psi-\delta^{i}_{k}\partial_{j}\psi\notag\\
&+&\partial_{j}\partial_{k}\partial^{_{i}}E.
\end{eqnarray}

\subsection{evolution equations for density perturbation and velocity perturbations}
In this section, we will given the derivation process of perturbed
energy-momentum equations. From the formula
\begin{eqnarray}
\delta\nabla_{\mu}T^{\mu0}&=&\delta g^{\mu\sigma}\overline{\nabla_{\mu}
T^{0}_{\sigma}}+\bar{g}^{\mu\sigma}\delta\nabla_{\mu}
T^{0}_{\sigma},\notag \\
\delta\nabla_{\mu}T^{\mu i}&=&\delta g^{\mu\sigma}\overline{\nabla_{\mu}
T^{i}_{\sigma}}+\bar{g}^{\mu\sigma}\delta\nabla_{\mu}
T^{i}_{\sigma},
\end{eqnarray}
and
\begin{eqnarray}
\nabla_{\mu}T^{\nu}_{\sigma}&=&T^{\nu}_{\sigma,\mu}+\Gamma^{\nu}_{\rho\mu}T^{\rho}_{\sigma}-\Gamma^{\rho}_{\sigma\mu}T^{\nu}_{\rho}, \notag\\
\overline{\nabla_{\mu}T^{\nu}_{\sigma}}&=&\bar{T}^{\nu}_{\sigma,\mu}+\bar{\Gamma}^{\nu}_{\rho\mu}\bar{T}^{\rho}_{\sigma}-\bar{\Gamma}^{\rho}_{\sigma\mu}\bar{T}^{\nu}_{\rho},
\end{eqnarray}
namely
\begin{eqnarray}
\overline{\nabla_{0}T^{0}_{0}}&=&-\bar{\rho}^{\prime},
~~~~~~~~\delta\nabla_{0}T^{0}_{0}=-\delta\rho^{\prime},\notag\\
\overline{\nabla_{0}T^{0}_{i}}&=&0,~~~~~~~~\overline{\nabla_{i}T^{0}_{j}}=\mathcal{H}(\bar{\rho}+\overline{p_{eff}})\delta_{ij},\notag\\
\overline{\nabla_{i}T^{0}_{0}}&=&0,~~~~~~\delta\nabla_{i}T^{0}_{0}=-\mathcal{H}(\bar{\rho}+\overline{p_{eff}})(2v_{i}+B_{i}),\notag\\
\delta\nabla_{0}T^{0}_{i}&=&[(\bar{\rho}+\overline{p_{eff}})(v_{i}+B_{i})]^{\prime}\notag\\
&&+(\bar{\rho}+\overline{p_{eff}})(\partial_{i}\phi+\mathcal{H}\partial_{i}B),\notag\\
\delta\nabla_{i}T^{0}_{j}&=&(\bar{\rho}+\overline{p_{eff}})(\partial_{i}\partial_{j}v+\partial_{i}\partial_{j}B)+\mathcal{H}(\delta\rho+\delta
p_{eff})\delta_{ij}\notag\\
&&-(\bar{\rho}+\overline{p_{eff}})[\psi^{\prime}+2\mathcal{H}(\psi+\phi)]\delta_{ij}\notag\\
&&+(\bar{\rho}+\overline{p_{eff}})\nabla^{2}(E^{\prime}+2\mathcal{H}E-B),
\end{eqnarray}
we obtain the following
perturbed energy-momentum equation
\begin{eqnarray}
&&\delta\nabla_{\mu}T^{\mu0}=\frac{1}{a^{2}}\{\delta\rho^{\prime}+3\mathcal{H}(\delta\rho+\delta
p_{eff})-3(\bar{\rho}+\overline{p_{eff}})\psi^{\prime}\notag\\
&&+(\bar{\rho}+\overline{p_{eff}})\nabla^{2}(v+E^{\prime})-2\phi[\bar{\rho}^{\prime}+3\mathcal{H}(\bar{\rho}+\overline{p_{eff}})]\}.
\end{eqnarray}
And in the same way, make use of the following results
\begin{eqnarray}
\overline{\nabla_{0}T^{i}_{0}}&=&0,~~~~~~\overline{\nabla_{j}T^{i}_{0}}=-\mathcal{H}(\bar{\rho}+\overline{p_{eff}})\delta^{i}_{j},\notag\\
\delta\nabla_{0}T^{i}_{0}&=&-[(\bar{\rho}+\overline{p_{eff}})v^{i}]^{\prime}\notag\\
&&-(\bar{\rho}+\overline{p_{eff}})\partial^{i}(\phi+B+\mathcal{H}B),\notag\\
\overline{\nabla_{0}T^{i}_{k}}&=&p^{\prime}_{eff}\delta^{i}_{k},~\delta\nabla_{0}T^{i}_{k}=\delta{p^{\prime}_{eff}}\delta^{i}_{k},\notag\\
\overline{\nabla_{j}T^{i}_{k}}&=&\partial_{j}(\overline{p_{eff}}\delta^{i}_{k}),\notag\\
\delta\nabla_{j}T^{i}_{k}&=&\mathcal{H}(\bar{\rho}+\overline{p_{eff}})[\delta^{i}_{j}\partial_{k}v+\delta^{i}_{j}\partial_{k}B+\delta_{kj}\partial^{i}v]\notag\\
&&+\partial_{j}(\delta
p_{eff}\delta_{k}^{i}),\notag\\
\delta\nabla_{j}T^{i}_{0}&=&-(\bar{\rho}+\overline{p_{eff}})(-\psi^{\prime}\delta^{i}_{j}+\partial^{i}\partial_{j}E^{\prime})\notag\\
&&-\mathcal{H}\delta^{i}_{j}(\delta\rho+\delta
p_{eff})-(\bar{\rho}+\overline{p_{eff}})\partial_{j}\partial^{i}v,
\end{eqnarray}
we also have the following equation
\begin{eqnarray}
\delta\nabla_{\mu}T^{\mu
i}&=&\frac{1}{a^{2}}\partial^{i}\{[(\bar{\rho}+\overline{p_{eff}})(v+B)]^{\prime}+(\bar{\rho}+\overline{p_{eff}})\phi
\notag\\
&+&4\mathcal{H}(\bar{\rho}+\overline{p_{eff}})(v+B)+\delta
p_{eff}\notag\\
&-&[\bar{\rho}^{\prime}
+3\mathcal{H}(\bar{\rho}+\overline{p_{eff}})]B\}.
\end{eqnarray}
If the fluid is conservation, i.e.
$\bar{\rho}^{\prime}+3\mathcal{H}(\bar{\rho}+\overline{p_{eff}})=0$,
the above perturbed equations can be rewritten as
\begin{eqnarray}
\delta\nabla_{\mu}T^{\mu0}&=&\frac{1}{a^{2}}\{\delta\rho^{\prime}+3\mathcal{H}(\delta\rho+\delta
p_{eff})-3(\bar{\rho}+\overline{p_{eff}})\psi^{\prime}\notag\\
&+&(\bar{\rho}+\overline{p_{eff}})\nabla^{2}(v+E^{\prime})\},
\end{eqnarray}
\begin{eqnarray}
\delta\nabla_{\mu}T^{\mu
i}&=&\frac{1}{a^{2}}\partial^{i}\{[(\bar{\rho}+\overline{p_{eff}})(v+B)]^{\prime}+(\bar{\rho}+\overline{p_{eff}})\phi\notag\\
&+&4\mathcal{H}(\bar{\rho}+\overline{p_{eff}})(v+B)+\delta p_{eff}
\},
\end{eqnarray}
where
$\overline{p_{eff}}=\bar{p}-\frac{3}{a}\mathcal{H}\zeta$,
\begin{eqnarray}
&&\delta p_{eff}=\delta p-\delta\zeta\overline{(\nabla_{\gamma}u^{\gamma})}-\zeta(\delta\nabla_{\gamma}u^{\gamma})\notag\\
&&=\delta
p-\frac{3\mathcal{H}}{a}\delta\zeta-\frac{\zeta}{a}[\nabla^{2}(v+E^{\prime})-(3\psi^{\prime}+3\mathcal{H}\phi)].
\end{eqnarray}
To solve the above equations or make the complete, we need the
relations between $\delta p$ and $\delta \rho$. The sound sound
speed $c_{s,eff}^{2}$ of a fluid or scalar field, is the propagation
speed of pressure fluctuation in the rest frame
\begin{equation}
c_{s,eff}^{2}=\frac{\delta p_{eff}}{\delta \rho}|_{rf},
\end{equation}
where '$|_{rf}$' denotes the rest frame. For scalar field $\phi$,
the rest frame is defined as the hypersurfaces $\delta\phi=0$, i.e.
$\phi=constant$. So, one has $\delta V=0$ and
$\delta\rho_{\phi}=\delta(\frac{1}{2}a^{-2}\phi^{\prime2}+V)=a^{-2}\phi^{\prime}\delta\phi^{\prime}=\delta
p_{\phi}$. Thus the sound speed of scalar field equals to the speed
of light, is independent the form of $V(\phi)$
\begin{equation}
\delta\phi|_{rf}=0\Rightarrow c_{s\phi}^{2}=1.
\end{equation}
The "adiabatic sound speed" for any medium is defined as
\begin{equation}
c_{a,eff}^{2}=\frac{p_{eff}^{\prime}}{\rho^{\prime}}=w_{eff}+\frac{w_{eff}^{\prime}}{\rho^{\prime}/\rho}.
\end{equation}
The rest frame (the
zero momentum gauge or comoving orthogonal gauge) is the comoving
$(v|_{rf}=0)$ orthogonal $(B|_{rf})=0)$ frame, so that
\begin{equation}
T_{0}^{i}|rf=0=T_{i}^{0}|rf.
\end{equation}
We make a gauge transformation,
$x^{\mu}\rightarrow
x^{\mu}+(\delta\tau, \partial^{i}\delta x)$, from
the rest frame gauge to a general gauge
\begin{equation}
v+B=(v+B)|_{rf}+\delta\tau,~~~\delta p=\delta
p|_{rf}-p^{\prime}\delta\tau,~~~\delta\rho=\delta\rho|_{rf}-\rho^{\prime}\delta\tau.
\end{equation}
Thus, one has $\delta\tau=v+B$ and
\begin{eqnarray}
\delta p_{eff}&=&\delta p_{eff}|_{rf}-p_{eff}^{\prime}\delta\tau\notag\\
&=&c_{s,eff}^{2}\delta\rho+\delta\rho_{nad},
\end{eqnarray}
where
$\delta\rho_{nad}=(c_{s,eff}^{2}-c_{a,eff}^{2})[\delta\rho+\rho^{\prime}(v+B)]$
is the intrinsic non-adiabatic perturbation in the fluid. When the
fluid is conservation, i.e.
$\bar{\rho}^{\prime}=-3\mathcal{H}(\bar{\rho}+\overline{p_{eff}})$.
By using the relation $\theta=-k^{2}(v+B)$ in Fourier space, one has
\begin{eqnarray}
&&\delta p_{eff}=c_{s,eff}^{2}\delta\rho+(c_{s,eff}^{2}-c_{a,eff}^{2})\rho^{\prime}(v+B)\notag\\
&&=c_{s,eff}^{2}\delta\rho+(c_{s,eff}^{2}-c_{a,eff}^{2})[3\mathcal{H}(\bar{\rho}+\overline{p_{eff}})]\frac{\theta}{k^{2}}.
\end{eqnarray}
We define the density contrast $\delta=\delta\rho/\bar{\rho}$, then one has the evolution equations for density perturbation and velocity perturbation for a generic conservation fluid are
\begin{eqnarray}
&&\delta^{\prime}+3\mathcal{H}(c_{s,eff}^{2}-w_{eff})\delta
+(1+w_{eff})(\theta-3\psi^{\prime})=0,\notag\\
\end{eqnarray}
\begin{equation}
\theta^{\prime}+\mathcal{H}(1-3c_{s,eff}^{2})\theta-\frac{c_{s,eff}^{2}k^{2}\delta}{1+w_{eff}}-k^{2}\phi=0.
\end{equation}
In the synchronous gauges, one has
\begin{eqnarray}
\phi&=&\beta^{\prime\prime}+\frac{a^{\prime}}{a}\beta^{\prime},\\\
\psi&=&-\frac{h}{6}-\frac{1}{3}\nabla^{2}\beta-\frac{a^{\prime}}{a}\beta^{\prime}.
\end{eqnarray}
Therefore, $k^{2}\phi=0$, ~$-3\psi^{\prime}=\frac{h^{\prime}}{2}$,
finally, we has the following evolution equations for density
perturbation and velocity perturbation
\begin{eqnarray}
\delta^{\prime}&=&-(1+w_{eff})(\theta+\frac{h^{\prime}}{2})-3\mathcal{H}(c_{s,eff}^{2}-w_{eff})\delta
,\label{eq:continue}\\
\theta^{\prime}&=&-\mathcal{H}(1-3c_{s,eff}^{2})\theta+\frac{c_{s,eff}^{2}k^{2}\delta}{1+w_{eff}}\label{eq:euler}.
\end{eqnarray}
Following the formalism for a generalized dark matter \cite{ref:Hu98}, one can recast Eqs. (\ref{eq:continue}), and (\ref{eq:euler}) into
\begin{eqnarray}
\delta^{\prime}&=&-(1+w_{eff})(\theta+\frac{h^{\prime}}{2})+\frac{w_{eff}^{\prime}}{1+w_{eff}}\delta\notag\\
&-&3\mathcal{H}(c_{s,eff}^{2}-c_{a,eff}^{2})[\delta+3\mathcal{H}(1+w_{eff})\frac{\theta}{k^{2}}]
,\\\
\theta^{\prime}&=&-\mathcal{H}(1-3c_{s,eff}^{2})\theta+\frac{c_{s,eff}^{2}k^{2}\delta}{1+w_{eff}},
\end{eqnarray}
where
\begin{eqnarray}
w_{eff}&=&-\frac{B_{s}}{B_{s}+(1-B_{s})a^{-3(1+\alpha)}}-\sqrt{3}\zeta_{0},\\\
c_{a,eff}^{2}&=&w_{eff}-\frac{w_{eff}^{\prime}}{3\mathcal{H}(1+w_{eff})},\\\
c_{s,eff}^{2}&=&c_{s}^{2}-\frac{\sqrt{3}}{2}\zeta_{0}-\frac{\zeta_{0}}{\sqrt{3}\mathcal{H}\delta}(\theta+\frac{h^{\prime}}{2}),\\\
c_{s,eff}^{2}&-&c_{a,eff}^{2}=\frac{w_{eff}\Gamma_{nad,eff}}{\delta^{rest}},\\\
\Gamma_{nad,eff}&=&\frac{\delta
p_{nad}}{p_{eff}},\\\
\delta^{rest}&=&\delta+3\mathcal{H}(1+w)\frac{\theta}{k^{2}}.
\end{eqnarray}

\section{Cosmological constraints From Data Sets: SNLS3, BAO, Planck And HST}
In this section, we apply the Markov Chain Monte Carlo method to
investigate the observational constraint on viscous
generalized Ghapylin gas model which included bulk viscous perturbation to obtaining the parameters space.
The MCMC method is based on the publicly available cosmoMC package
\cite{ref:MCMC}, which has been modified to include the dark fluid
perturbation in the CAMB \cite{ref:CAMB} code which is used to
calculate the theoretical CMB power spectrum. To get the converged
results, in MCMC calculation we stop sampling by checking the worst
e-values [the variance(mean)/mean(variance) of 1/2 chains] $R-1$ of
the order $0.01$. In the following calculations, we take the total
likelihood  $\mathcal{L} \propto e^{-\chi^{2}/2}$  to be the product
of the separate likelihoods of SNLS3, BAO, Planck and HST. Then the $\chi^{2}$
is given as
\begin{equation}
\chi^{2}=\chi^{2}_{SNLS3}+\chi^{2}_{BAO}+\chi^{2}_{Planck}+\chi^{2}_{HST},
\end{equation}
with the following 8-dimensional parameter space:
\begin{eqnarray}
&&P\equiv(\omega_b, 100\theta_{MC}, \tau, \alpha, B_{s},
\zeta_{0}, n_s, \log[10^{10}A_s]).
\end{eqnarray}
The pivot scale of the initial scalar power spectrum
$k_{s0}=0.05\text{Mpc}^{-1}$ is used and the priors to model
parameters is taken as follows: the physical baryon density
$\omega_{b}(=\Omega_{b}h^{2})\in [0.005,0.1]$;  the ratio of the
sound horizon and angular diameter distance $100\theta_{MC}\in[0.5,10]$;
the optical depth $\tau\in[0.01,0.8]$;  the model parameters
$\alpha\in[0,0.1]$, $B_{s}\in[0,1]$ and  $\zeta_{0}\in[0,0.01]$;
 the scalar spectral index $n_{s}\in[0.5,1.5]$, and logarithm of the
amplitude of the initial power spectrum $\log[10^{10}A_{s}]\in[2.7,
4]$. In addition, the hard coded prior on the comic age
$10\text{Gyr}<t_{0}<\text{20Gyr}$ is imposed. Also, the weak
Gaussian prior on the physical baryon density
$\omega_{b}=0.022\pm0.002$ \cite{ref:bbn} from big bang
nucleosynthesis and new Hubble constant
$H_{0}=74.2\pm3.6\text{kms}^{-1}\text{Mpc}^{-1}$ \cite{ref:hubble}
are adopted. Notice that the current dimensionless energy density of
VGCG $\Omega_{VGCG}$ is not included in the model parameter space $P$,
because it is a derived parameter in a spatially flat ($k=0$) FRW
universe. To study the evolutions of the perturbation, we should fix the background evolution. To realize that, we use the cosmic observations from the type Ia supernovae SNLS3, cosmic microwave background radiation from recently released Planck, baryon acoustic oscillation from Sloan Digital Sky Survey and the WiggleZ data points and High -redshift SN observations from Hubble Space Telescope. For the detailed description, please see Refs.\cite{xu1}\cite{xu2}.
\begingroup
\squeezetable
\begin{center}
\begin{table}
\begin{tabular}{cc}
\hline\hline
Model Parameters & Mean value with errors \\ \hline
$\Omega_b h^2$ & $    0.0222_{-    0.000303-    0.000590-    0.000781}^{+    0.000302+    0.000603+    0.000802}$\\

$100\theta_{MC}$ & $    1.051_{-    0.000558-    0.00110-    0.00143}^{+    0.000553+    0.00109+    0.00144}$\\

$\tau$ & $    0.0854_{-    0.01354-    0.0238-    0.0309}^{+    0.0121+    0.0259+    0.0347}$\\

$\alpha$ & $    0.192_{-    0.134-    0.192-    0.192}^{+    0.0835+    0.195+    0.292}$\\

$Bs$ & $    0.808_{-    0.0334-    0.0624-    0.0710}^{+    0.0328+    0.0629+    0.0807}$\\

$\zeta_0$ & $    0.0000138_{-    0.0000105-    0.0000138-    0.0000138}^{+    0.00000614+    0.0000145+    0.0000212}$\\

$n_s$ & $    0.964_{-    0.00710-    0.0138-    0.0181}^{+    0.00714+    0.0141+    0.0185}$\\

${\rm{log}}(10^{10} A_s)$ & $    3.0820_{-    0.0262-    0.0470-    0.0615}^{+    0.0238+    0.0502+    0.0660}$\\

$\Omega_{VGcG}$ & $    0.955_{-    0.00173-    0.00322-    0.00413}^{+    0.00172+    0.00331+    0.00422}$\\

$\Omega_b$ & $    0.0453_{-    0.00171-    0.00331-    0.00422}^{+    0.00173+    0.00322+    0.00413}$\\

$z_{re}$ & $   10.626_{-    1.0813-    2.159-    2.900}^{+    1.101+    2.172+    2.834}$\\

$H_0$ & $   71.0621_{-    1.349-    2.357-    3.0527}^{+    1.202+    2.504+    3.287}$\\

${\rm{Age}}/{\rm{Gyr}}$ & $   13.723_{-    0.0397-    0.0791-    0.106}^{+    0.0395+    0.0797+    0.103}$\\
\hline\hline
\end{tabular}
\caption{The mean values of model parameters with $1\sigma$,
$2\sigma$ and $3\sigma$ errors from the combination
SNLS3+BAO+Planck+HST.}\label{tab:results}
\end{table}
\end{center}
\endgroup

\begin{center}
\begin{figure}[tbh]
\includegraphics[width=9cm]{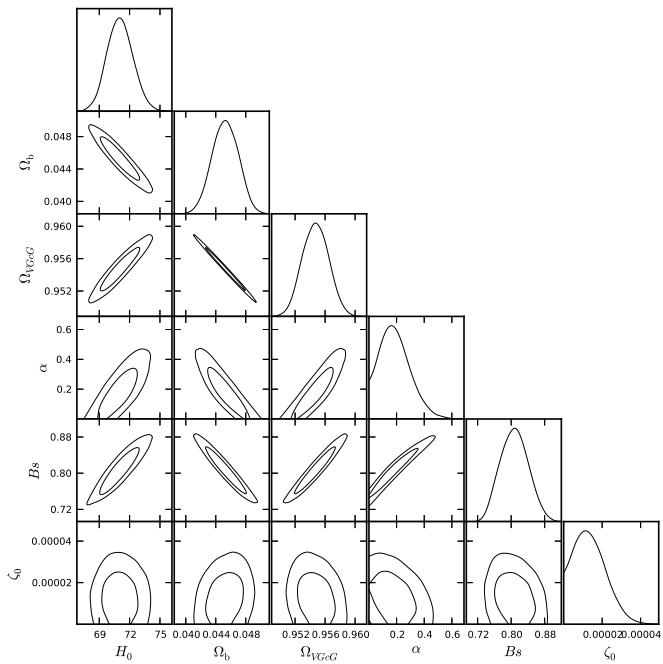}
\caption{The 1D marginalized distribution on individual parameter
and 2D contours  with $68\%$ C.L., $95\%$ C.L., and $99\%$ C.L. by
using SNLS3+BAO+Planck+HST data points.}\label{fig:contour}
\end{figure}
\end{center}

The best fitting values of the cosmological parameters and  the mean
values of model parameters with $1\sigma$, $2\sigma$ and $3\sigma$
regions in VGCG model from the combination SNLS3+BAO+Planck+HST are listed in
Table \ref{tab:results}. Correspondingly, the contour plots are
shown in Figure \ref{fig:contour}. We find that the minimum
$\chi^{2}$ is $\chi^{2}_{min}=5115.878$. From Table
\ref{tab:results} and Figure \ref{fig:contour}, we obtain the constraint on the bulk viscosity coefficient: $\zeta_0=0.0000138_{-    0.0000105-    0.0000138-    0.0000138}^{+    0.00000614+    0.0000145+    0.0000212}$ in $1, 2, 3\sigma$ regions respectively, it is obvious that
we obtain a tighter constraint than our previous results in \cite{15} due to the bulk viscosity perturbation
is included.
From \cite{15}, we know that the value of bulk viscosity impacts the CMB
power spectrum on its height of the peak sensitively. Since
the parameter $\zeta_{0}$ is related to the dimensionless density parameter
of effective cold dark matter $\Omega_{c0}$, decreasing the
values of $\zeta_{0}$ is equivalent to increase the value of effective
dimensionless energy density of cold dark matter, so the smaller bulk viscosity $\zeta_{0}$ will
make the equality of matter and radiation earlier, therefore
the sound horizon is decreased, this can be embodied in the CMB anisotropic power spectra by showing the first peak is depressed as observed in the figure 2 in \cite{15}.

\begin{center}
\begin{figure}[tbh]
\includegraphics[width=8.5cm]{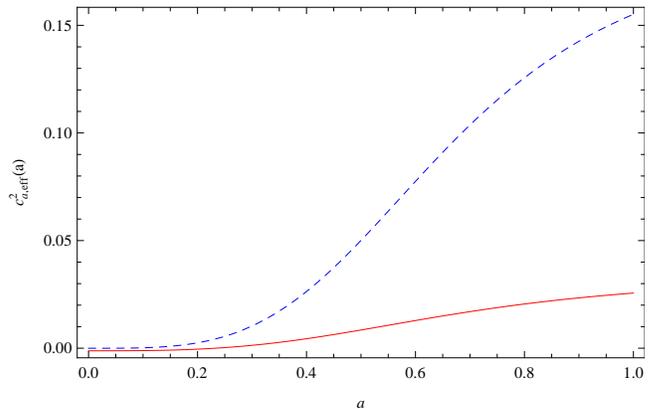}
\caption{The evolution of $c^{2}_{a,eff}$ with respect to scale factor $a$. The solid curve
correspond to VGCG2 model (including bulk viscosity perturbation); the dashed curve
correspond to VGCG1 model (not considering bulk viscosity perturbation).
}\label{fig:cs}
\end{figure}
\end{center}

\begin{center}
\begin{figure}[tbh]
\includegraphics[width=8.5cm]{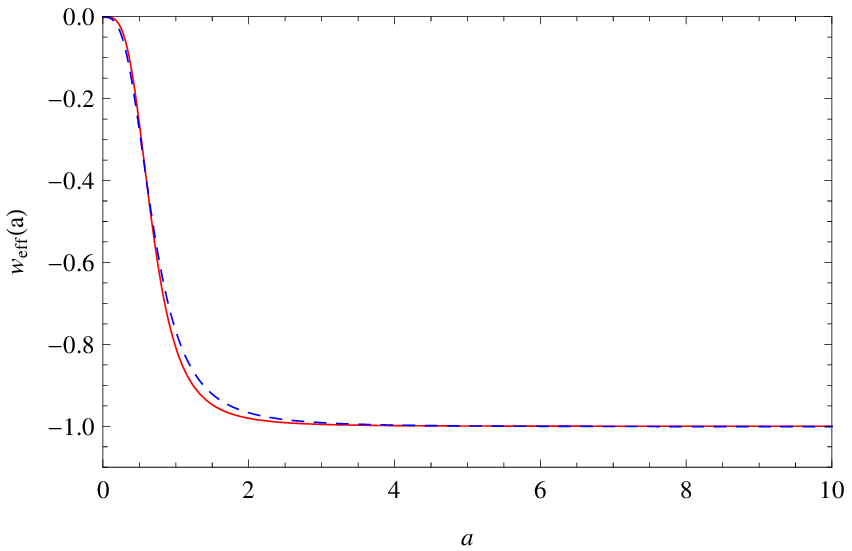}
\includegraphics[width=8.5cm]{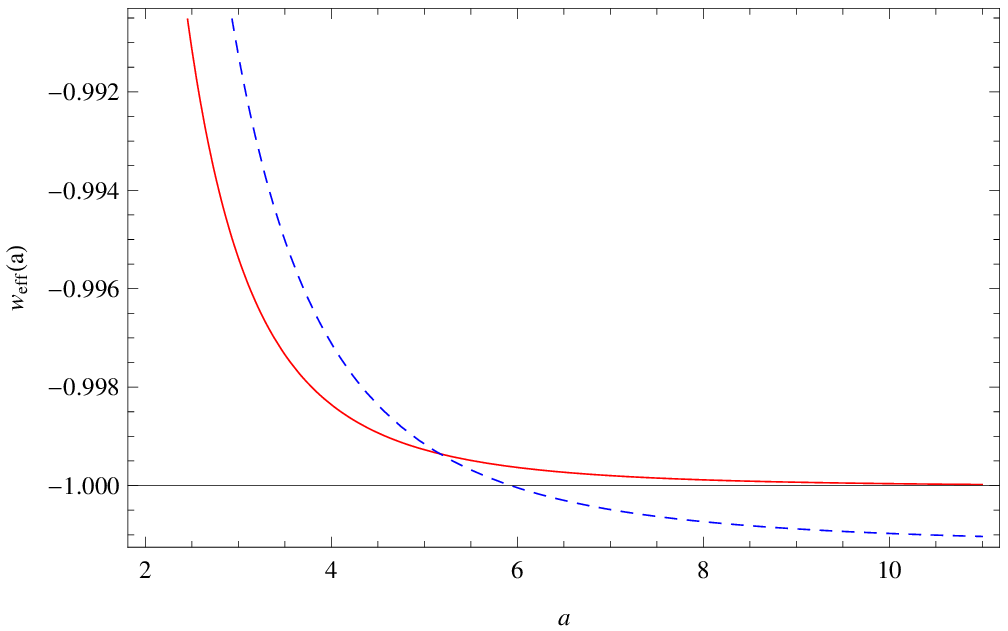}
\caption{The evolution of $w_{eff}$ with respect to scale factor $a$. The solid curve
correspond to VGCG2 model (including bulk viscosity perturbation); the dashed curve
correspond to VGCG1 model (not considering bulk viscosity perturbation).
}\label{fig:wa}
\end{figure}
\end{center}

To show the effect of bulk viscosity perturbation to the efficient state parameter $w_{eff}$ and the efficient adiabatic sound speed $c_{a,eff}^{2}$, we plot the
the evolution curves of $c_{a,eff}^{2}$ and $w_{eff}$ with respect to scale
factor $a$ in Figure \ref{fig:cs} and Figure \ref{fig:wa} respectively, which corresponding to VGCG1 model (not considering bulk viscosity perturbation) and VGCG2 model (including bulk viscosity perturbation).
From Figure \ref{fig:cs}, one can conclude that VGCG2 model provides a more smaller efficient adiabatic sound speed (which approximately equal to zero) than VGCG1 model. It is well known that almost zero adiabatic sound speed which being characterized by the perturbation of density contrast is important for large scale structure
formation. So, VGCG2 model make it possible to form large scale structures in our universe. From the upper panel of Figure \ref{fig:wa}, one can see that the two VGCG models behave like cold dark matter with almost zero EoS at early epoch ($a < 0.2$), and behave like dark energy with EoS $w_{eff} < 0$ at late time, which pushes the universe into an accelerated phase. Furthermore, from the under panel of Figure \ref{fig:wa}, which enlarged the upper panel (from $a=2$ to the end ), we can conclude that VGCG1 model behaves like quintessence ($w_{eff} > -1$) at present, behaves like phantom ($w_{eff} < -1$) in the distant future. However, unlike  VGCG1 model,  VGCG2 model behaves like quintessence at present and in the distant future, which will avoid our universe to be terminated by a cosmic doomsday. Therefore, it is more necessary and reasonable to include the perturbation of bulk viscosity when we study of cosmic  evolution. In conclusion, VGCG2 model (including bulk viscosity perturbation) being proposed here is a more competitive model than the one we studied previously.

\section{Discussion And Conclusion}
In this paper, we have revisited the viscous generalized
Chaplygin gas (VGCG) model by including perturbation of bulk viscosity. We derived the cosmological evolution equations for density perturbation and velocity perturbation.
By using MCMC method with the combination of SNLS3, BAO, HST and recently released Planck
data points, we obtained tighter
constraints as shown in the forth section of this paper.  Since the parameter $\zeta_{0}$ is related to the dimensionless
density parameter of effective cold dark matter, decreasing
the values of $\zeta_{0}$ is equivalent to increase the value
of effective dimensionless energy density of cold dark matter,
then it will make the equality of matter and radiation
earlier, therefore the sound horizon is decreased. So we
predict that the more smaller bulk viscosity coefficient parameter
$\zeta_0=0.0000138_{-    0.0000105-    0.0000138-    0.0000138}^{+    0.00000614+    0.0000145+    0.0000212}$ in $1, 2, 3\sigma$ regions respectively will depress the peak of the  decreases CMB $C^{TT}_{l}$
power spectrum on its height. From Figure \ref{fig:cs}, one can conclude that VGCG2 model provides a more smaller efficient adiabatic sound speed which is important for large scale structure formation than VGCG1 model. So, VGCG2 model make it possible to form large scale structures in our universe. From Figure \ref{fig:wa}, one can see that the two VGCG models behave like cold dark matter with almost zero EoS at early epoch ($a < 0.2$), and behave like dark energy with EoS ($w_{eff} < 0$) at late time, which pushes the universe into an accelerated phase. Furthermore, we can see that VGCG1 model behaves like quintessence ($w_{eff} > -1$) at present, behaves like phantom ($w_{eff} < -1$)in the distant future. However, unlike the VGCG1 model, the VGCG2 model behaves like quintessence at present and in the distant future, which will avoid our universe to be terminated by a cosmic doomsday. Therefore, it is more reasonable to include perturbation of bulk viscosity when we study of cosmic  evolution. Because of the almost zero sound speed and almost negative one state parameter (in the distant future), we come to a conculsion that the viscous generalized Chaplygin gas model which including bulk viscosity perturbation is a competitive replacement of $\Lambda$CDM model.

\section{Acknowledgements}
L. Xu's work is supported in part by NSFC under the Grants No. 11275035 and "the Fundamental Research Funds for the Central Universities" under the Grants No. DUT13LK01.

\end{document}